# Observation of high-temperature superconductivity in the high-pressure tetragonal phase of La$_2$PrNi$_2$O$_{7-\delta}$

G. Wang[1,2#], N. N. Wang[1,2#*], Y. X. Wang[1,2#], L. F. Shi[1,2], X. L. Shen[3], J. Hou[1,2], H. M. Ma[3], P. T. Yang[1,2], Z. Y. Liu[1,2], H. Zhang[1,2], X. L. Dong[1,2], J. P. Sun[1,2], B. S. Wang[1,2], K. Jiang[1,2*], J. P. Hu[1,2], Y. Uwatoko[3], and J.-G. Cheng[1,2*]

[1]*Beijing National Laboratory for Condensed Matter Physics and Institute of Physics, Chinese Academy of Sciences, Beijing 100190, China*

[2]*School of Physical Sciences, University of Chinese Academy of Sciences, Beijing 100190, China*

[3]*Institute for Solid State Physics, University of Tokyo, Kashiwa, Chiba 277-8581, Japan*

# These authors contribute equally to this work.

*Corresponding authors: nnwang@iphy.ac.cn; jiangkun@iphy.ac.cn; jgcheng@iphy.ac.cn

## Abstract

The recent discovery of high-temperature superconductivity in the Ruddlesden-Popper phase La$_3$Ni$_2$O$_7$ under high pressure marks a significant breakthrough in the field of 3$d$ transition-metal oxide superconductors. For an emerging novel class of high-$T_c$ superconductors, it is crucial to find more analogous superconducting materials with a dedicated effort toward broadening the scope of nickelate superconductors. Here, we report on the observation of high-$T_c$ superconductivity in the high-pressure tetragonal *I4/mmm* phase of La$_2$PrNi$_2$O$_{7-\delta}$ above ~10 GPa, which is distinct from the reported orthorhombic *Fmmm* phase of La$_3$Ni$_2$O$_7$ above 14 GPa. For La$_2$PrNi$_2$O$_{7-\delta}$, the onset and the zero-resistance temperatures of superconductivity reach $T_c^{onset}$ = 78.2 K and $T_c^{zero}$ = 40 K at 15 GPa. This superconducting phase shares the similar structural symmetry as many cuprate superconductors, providing a fresh platform to investigate underlying mechanisms of nickelate superconductors.

# Introduction

Following the discovery of high-temperature superconductivity in cuprates[1-3], numerous research endeavors have been dedicated to exploring more superconducting material classes with high $T_c$ and to elucidating the underlying mechanisms. To date, high-$T_c$ superconductivity has been realized in a limited number of material families, including cuprates[2-4] and iron-based superconductors[5,6], whose microscopic mechanism remains unresolved. The recent report of high-$T_c$ superconductivity in the $La_3Ni_2O_7$ crystal[7-10] at pressures above 14 GPa stands out as a conspicuous breakthrough in the realm of $3d$ transition-metal oxides, which has immediately emerged as a central subject of many follow-up studies[7-20]. The appearance of superconductivity with the highest $T_c \approx 80$ K in the pressurized $La_3Ni_2O_{7-\delta}$ was reported to have a close correlation with the structure transformation from the orthorhombic *Amam* (No. 63) to the orthorhombic *Fmmm* (No. 69) under high pressures around 10-15 GPa. This phenomenon suggests that the distinct crystal and electronic structures induced by external pressure play a decisive role in the generation of superconductivity in $La_3Ni_2O_7$. In addition, first-principle calculations suggest that the interplay between the half-filling $3d_{z^2}$ and the quarter-filling $3d_{x^2-y^2}$ orbitals is a pivotal factor contributing to the emergence of high-temperature superconductivity in $La_3Ni_2O_7$ [13-17,21-27].

As an emerging novel class of high-$T_c$ superconductors, it is highly desirable to find more analog superconducting materials with similar structural units in order to broaden the scope of nickelate superconductors. One of the common approaches is to replace $La^{3+}$ with isovalent smaller-size rare-earth $R^{3+}$ ions or substitute it with heterovalent cations. Recent theoretical investigations on the $R_3Ni_2O_7$ ($R$ = rare earth) presented distinct viewpoints regarding the influence of chemical precompression induced by the substitution of $R^{3+}$ with smaller ionic radius[28-30]. In addition, the experimental verifications are still lacking due to the fact that, among $R_3Ni_2O_7$, only the pristine $La_3Ni_2O_7$ and partially substituted $La_{3-x}R_xNi_2O_7$ can be obtained in experiments. From the perspective of materials synthesis, partial replacements of $La^{3+}$ with smaller $Pr^{3+}$ ions introduce chemical pressure, which is expected to reduce the physical pressure required to induce structural transition and superconductivity. We are thus motivated to investigate the pressure effect on the structural and transport properties of Pr-doped $La_{3-x}Pr_xNi_2O_7$ polycrystalline samples.

In this work, we synthesized a series of $La_{3-x}Pr_xNi_2O_7$ ($x$ = 0, 0.1, 0.3, 1) polycrystalline samples using the sol-gel method, and systematically investigated the evolution of their structure, magnetic, and electrical transport properties as a function of doping level at ambient pressure. Our results confirm that all these samples maintain the orthorhombic structure with the space group *Amam* (No.63) at ambient conditions. In addition, we observe a consistent negative correlation between the lattice parameters and the doping level ($x$) in $La_{3-x}Pr_xNi_2O_7$, following well with the Vegard's law. Then we selected the sample $La_2PrNi_2O_{7-\delta}$ with the highest doping level, and performed detailed high-pressure measurements on the crystal structural and electrical transport properties. Interestingly, we find that the chemically precompressed $La_2PrNi_2O_{7-\delta}$ transforms from the orthorhombic *Amam* space group to the tetragonal *I4/mmm* space group at about 10-11 GPa, accompanied by the appearance of high-temperature superconductivity. The onset and the zero-resistance temperatures of superconductivity reach $T_c^{onset}$ = 78.2 K and $T_c^{zero}$ = 40 K at 15 GPa. Different from the orthorhombic *Fmmm* phase observed in $La_3Ni_2O_7$[7], the discovery of superconductivity in the high-pressure tetragonal phase of $La_2PrNi_2O_{7-\delta}$ expands the family of nickelate superconductors and provides a fresh platform for investigating the underlying mechanisms at play.

**Methodology**

**Sample synthesis.** $La_{3-x}Pr_xNi_2O_7$ (x = 0, 0.1, 0.3, 1) samples were synthesized with the sol-gel method as described in previous studies[31,32]. Stoichiometric mixtures of $La_2O_3$, $Pr_6O_{11}$ and $Ni(NO_3)_2 \cdot 6H_2O$, all of them with purity of 99.99% from Alfa Aesar, were firstly dissolved in the deionized water with the addition of appropriate amount of citric acid and nitric acid, and stirred in a 90 °C water bath for approximately 4 hours. Then the obtained vibrant green nitrate gel was heat treated overnight at 800°C to remove excess organic matter. After that, the product was ground and pressed to pellets and post-sintered in air at 1100-1150 °C for 48 h.

**Sample characterizations.** The crystal structure of $La_{3-x}Pr_xNi_2O_7$ (x = 0, 0.1, 0.3, 1) at ambient pressure was determined by X-ray diffraction (XRD) collected through PANalytical X'Pert PRO with Cu $K_\alpha$ radiation. Temperature-dependent resistivity $\rho(T)$ and magnetic susceptibility $\chi(T)$ were measured by using the Quantum Design Physical Properties Measurement System (PPMS) and Magnetic Property Measurement System (MPMS), respectively.

**High-pressure measurements.** The high-pressure synchrotron XRD (HP-SXRD) of $La_2PrNi_2O_{7-\delta}$ was measured at the 4W2 beamline at the Beijing Synchrotron Radiation Facility (BSRF) with a wavelength of $\lambda$ = 0.6199 Å. Rietveld analysis of HP-SXRD data was performed with GSAS-II suite[33]. We employ the cubic anvil cell (CAC) to measure the $\rho(T)$ at different pressures up to 15 GPa by employing glycerol as the liquid pressure transmitting medium, which can afford an excellent hydrostatic pressure condition.

**DFT structures calculation.** We employ the Vienna ab initio simulation package (VASP) code[34] with the projector augmented wave (PAW) method[35] to do the density functional theory (DFT) calculation. The Perdew-Burke-Ernzerhof (PBE) exchange-correlation functional[36] is used in our calculations. The kinetic energy cutoff is set to be 500 eV for expanding the wave functions into a plane-wave basis and the energy convergence criterion is $10^{-8}$ eV. We replace each La in the LaO plane at the center of every $La_3Ni_2O_7$ layer with Pr to simulate 1/3 doping level. In principle, there are other possibilities for Pr to replace the La site in $La_3Ni_2O_7$, which shows less impact on the electronic structure of Ni. When considering the influence of pressure on crystal symmetry, atomic positions and lattice are fully relaxed until the atomic forces are less than 0.01 eV/Å. In band calculations, only the atomic positions are relaxed until the force is less than 0.01 eV/Å with the constrained experimental lattice constants. A $\Gamma$-centered 14×14×14 k-mesh is used for the primitive cell of HP phase and a $\Gamma$-centered 10×10×9 k-mesh is utilized for the primitive cell of low-pressure (LP) phase.

## Results and discussion

At first, we performed detailed characterizations on the structure, electrical transport, and magnetic properties of the series of $La_{3-x}Pr_xNi_2O_{7-\delta}$ (x = 0, 0.1, 0.3, 1) samples at ambient pressure. Figure 1(a) shows the powder XRD patterns of the synthesized $La_{3-x}Pr_xNi_2O_{7-\delta}$ (0.0⩽x⩽1.0) samples, which are confirmed to be single phase with the orthorhombic *Amam* (No.63) structure. The XRD data are refined via the Rietveld method using the FullProf suite software to extract structure information of these samples. As illustrated in Fig. S1, the refinements converged well for the entire $La_{3-x}Pr_xNi_2O_{7-\delta}$ (0.0⩽x⩽1.0) series. The obtained lattice parameters are displayed in Fig. 1(b) as a function of the Pr-content, *x*. As *x* increases in the $La_{3-x}Pr_xNi_2O_{7-\delta}$ series, the values of *a* and *b* exhibit minimal changes, while *c* decreases obviously resulting in a net reduction of unit-cell volume *V*, in line with the fact that $Pr^{3+}$ has a smaller ionic radius than $La^{3+}$, *i.e.* 1.126 Å versus 1.16 Å in the case of eight coordination[37]. The unit-cell parameters are *a* = 5.3920(2) Å, *b* = 5.4480(2) Å, *c* = 20.5311(9) Å, and *V* = 603.11(4) Å$^3$ for $La_3Ni_2O_{7-\delta}$, and *a* = 5.3724(3) Å, *b* = 5.4512(3) Å, *c* = 20.4073(13) Å, and *V* = 597.65(6) Å$^3$ for $La_2PrNi_2O_{7-\delta}$. This observation signifies that we have successfully introduced some chemical pressure in the $La_3Ni_2O_7$ lattice by replacing one-third of $La^{3+}$ with $Pr^{3+}$. Here, it is noted that we attempted to increase the $Pr^{3+}$ content in the precursor in order to introduce more chemical pressure in the $La_{3-x}Pr_xNi_2O_{7-\delta}$ lattice. However, this endeavor was unsuccessful, as the resultant product contains a significant amount of $La(Pr)Ni_2O_{4\pm\delta}$ impurities.

The temperature dependencies of $\rho(T)$ and $\chi(T)$ displayed in Fig. 1 (c) and (d) illustrated how the electrical transport and magnetic properties of $La_{3-x}Pr_xNi_2O_{7-\delta}$ evolve with the Pr substitution. The parent compound $La_3Ni_2O_{7-\delta}$ is a paramagnetic weak insulator, exhibiting a density-wave-like transition around 100-140 K, as evidenced by the anomalies in magnetic susceptibility and resistivity, in consistent with those reported in the literature[31,38-40]. As the Pr content increases, the magnetism susceptibility of $La_{3-x}Pr_xNi_2O_{7-\delta}$ progressively enhances and the low-temperature magnetic susceptibility of $La_2PrNi_2O_{7-\delta}$ increases by nearly two orders of magnitude compared with $La_3Ni_2O_{7-\delta}$. In addition, the temperature-dependent resistivity of $La_{3-x}Pr_xNi_2O_{7-\delta}$ gradually transfroms into an insulating behavior in the entire measured temperature range as the Pr content increases. However, the magnitude of the resistivity is still relatively low in comparison with the typical insulators. These observations indicate that the introduction of magnetic $Pr^{3+}$ imparts some local magnetic moments to the lattice, causing magnetic scattering during electron transport.

For $La_3Ni_2O_{7-\delta}$, the emergence of superconductivity under high pressure is closely associated with the structure phase transition from orthorhombic *Amam* (No.63) to orthorhombic *Fmmm* (No. 69) in the range of 10-15 GPa. It is thus highly interesting to examine whether the structural phase transition and high-temperature superconductivity would also appear at a lower pressure in the chemically precompressed $La_{3-x}Pr_xNi_2O_{7-\delta}$ samples. To this end, we chose to study the pressure effect on the $La_2PrNi_2O_{7-\delta}$ sample with the highest Pr content and proceeded to perform *in situ* HP-SXRD measurements. Figure 2(a) shows the XRD patterns of $La_2PrNi_2O_{7-\delta}$ collected at room temperature under various pressures from 3.2 to 56.0 GPa. The synchrotron XRD patterns acquired at pressures below 10.3 GPa can be fitted well with the orthorhombic structure with the space group *Amam*, as shown by the representative refinement at 3.2 GPa, Fig. 2(c). Upon further compressed to 11.2 GPa, several adjacent peaks merge together, such as the (020) and (200) peaks at $2\theta \sim 13.4°$, (135) and (315) peaks at $2\theta \sim 23.1°$, as shown in Fig. 2(b). This suggests the occurrence of a pressure-induced structural phase transition, most likely to a higher symmetry with a tetragonal structure. Subsequent structural analysis of the high-pressure phase revealed that its XRD patterns can be well described using the $Sr_3Ti_2O_7$-type structure model with the tetragonal *I4/mmm* space group (No. 139), as shown by the representative refinement at 11.1 GPa, Fig. 2(d). The observed structural transition in $La_2PrNi_2O_{7-\delta}$ is further supported by our DFT calculation. As shown in Table S1, DFT relaxation indicates that, for Pr-doped $La_2PrNi_2O_7$, the low-pressure phase tends to be the orthormhombic strcutre, whereas the high-pressure phase prefers the tetragonal structure, even at an external pressure of virtual 0 GPa. It is worth noting that such a structural phase transition from orthorhombic to tetragonal structure is distinct from that observed in $La_3Ni_2O_7$ where the high-pressure phase shares the same orthorhombic symmetry as the low-pressure phase. The lattice parameters extracted from the HP-SXRD patterns after Rietveld refinements are displayed in Fig. 2(e) and (f) as a function of pressure.

As can be seen, the lattice parameters decrease continuously with increasing pressure, but exhibit anisotropic compression behaviors. In the lower pressure range, lattice parameter $b$ decreases faster than $a$ and merges together with $a$ at 11.1 GPa, which is well consistent with the results seen from the HP-SXRD patterns. As the crystal structure transforms into a more symmetrical tetragonal structure, lattice parameter $a$ contracts by a factor of $\frac{1}{\sqrt{2}}$, leading to a 0.5 times reduction in the unit cell volume, $V$. The collected pressure-volume data in the whole pressure range can be fitted to the the third-order Birch-Murnaghan equation[41], which yields a bulk modulus $B_0$ = 142.7 GPa with $B_0'$ fixed at 3.5, as shown by the solid line in Fig. 2(f).

The above *in situ* high-pressure structural analysis surprisingly reveals that the chemically precompressed $La_2PrNi_2O_{7-\delta}$ sample exhibits a different, higher symmetry phase at pressures above 10-11 GPa compared to $La_3Ni_2O_7$ [7]. It is generally recognized that the crystal structure strongly influences the physical properties of condensed matter. The emergence of such a new tetragonal symmetry structure is expected to result in a distinct electronic structure in the high-pressure phase of $La_2PrNi_2O_{7-\delta}$, which, in turn, would directly impact its electrical transport properties. To explore the potential for superconductivity in the high-pressure tetragonal phase of $La_2PrNi_2O_{7-\delta}$, we performed high-pressure $\rho(T)$ measurements by using the cubic anvil cell under various hydrostatic pressures up to 15 GPa. Fig. 3(a) and (b) display the evolution of $\rho(T)$ for the polycrystalline $La_2PrNi_2O_{7-\delta}$ sample under various pressures. At 0 GPa, the $\rho(T)$ of $La_2PrNi_2O_{7-\delta}$ exhibits a typical insulating or semiconducting behavior throughout the entire temperature range, which is quite different from the weak insulating nature in undoped $La_3Ni_2O_7$[7,31,39,42]. As depicted in Fig. 3(a), the magnitude of $\rho(T)$ decreases monotonically in the whole temperature range but remains a semiconducting-like behavior as the pressure increases to 5 GPa. Upon further increasing pressure to 8 GPa, which approaches the phase transition pressure, a clear resistivity drop below 21.4 K is observed in $\rho(T)$ data, as shown in Fig. 3(b). This behavior becomes more pronounced as pressure increases, signaling the appearance of superconducting transition. At 12 GPa, $\rho(T)$ reaches zero resistance at 4.4 K with an onset temperature of $T_c^{onset}$ = 66.4 K. Upon further increasing pressure, the superconducting transition temperature $T_c^{zero}$ increases rapidly from 4.4 K at 12 GPa to 40 K at 15 GP and the onset temperature $T_c^{onset}$ reaches 78.2 K at 15 GPa. The occurrence of resistance anomaly in the $La_2PrNi_2O_{7-\delta}$ under high pressure is closely associated with its phase transition pressure which is similar to that observed in un-doped $La_3Ni_2O_7$ [7].

To further clarify the observed resistance drop is indeed associated with a superconducting transition, we performed $\rho(T)$ measurements at 15 GPa under different external magnetic fields. As displayed in Fig. 3(c), the resistivity below $T_c^{onset}$ is gradually suppressed by magnetic field, in line with the superconducting transition. The upper critical field $\mu_0H_{c2}(T_c)$ was determined using the criteria of 90% and 50% normal-state resistance at $T_c^{onset}$, all illustrated in Fig. 3(d). The zero-temperature-limit upper critical field $\mu_0H_{c2}(0)$ was estimated by the empirical Ginzburg-Landau equation as $\mu_0H_{c2}(0) = 103.3$ T and 50.8 T for $T_c^{90\%}$ and $T_c^{50\%}$, respectively.

To elucidate the electronic structure of the high-pressure tetragonal phase of $La_2PrNi_2O_{7-\delta}$ and to better understand the observed high-temperature superconductivity, we conducted DFT band calculations for $La_3Ni_2O_7$ and $La_2PrNi_2O_{7-\delta}$ at selected pressures. Note that the $Pr^{3+}$ has the same valance as the $La^{3+}$ and the Ni-O octahedra is elongated which is also the same as the situation of $La_3Ni_2O_7$. Consequently, the local crystal field environment is expected to be identical to $La_3Ni_2O_7$, with a $d^{7.5}$ filling of Ni and the splitting of two $d_{z2}$ orbitals due to the shared apical oxygen. The results of DFT band calculations are depicted in Fig. 4, which are consistent with the crystal field analysis. The splitting of two $d_{z2}$ orbitals of Ni results in a pure environment for $d_{x2-y2}$ orbitals of Ni near the fermi surface which is considered to be responsible for the high-$T_c$ superconductivity. As the pressure increases, the splitting is stronger. The difference of these two materials in the band structure is that the splitting is slightly stronger for $La_2PrNi_2O_7$. This can be attributed to the smaller radius of $Pr^{3+}$ compared to $La^{3+}$, inducing chemical stress in this material. In addition, in tetragonal phase of $La_2PrNi_2O_7$, the $d_{x2-y2}$ orbital of Ni exhibits stronger in-plane coupling to the $p$ orbitals of O to mediate superexchange couplings compared to orthorhombic phase of $La_3Ni_2O_7$. This is attributed to the fact that Ni form a square lattice in the tetragonal phase but a rhombohedral lattice in the orthorhombic phase. This should typically lead to higher optimal superconducting transition temperatures.

Figure 5 presents the constructed temperature-pressure ($T$-$P$) phase diagram of $La_2PrNi_2O_7$ polycrystalline samples. The $T_c$ values of undoped $La_3Ni_2O_7$ polycrystalline samples from Ref. [43] are also included for comparison. The phase diagram illustrates that high-temperature superconductivity of $La_2PrNi_2O_{7-\delta}$ emerges near the critical pressure of structure phase transition, which is similar to that observed in un-doped $La_3Ni_2O_7$ [43]. Upon further increasing pressure, the superconducting transition temperature $T_c^{onset}$ increases rapidly from 21.4 K at 8 GPa to 78.2 K at 15 GP and the superconducting transition $T_c^{zero}$ reaches 40 K at 15 GPa. It is noteworthy that the slope of $dT_c^{zero}/dP = 11.9$ K/GPa for $La_2PrNi_2O_{7-\delta}$ is much higher than that of 4.5 K/GPa for $La_3Ni_2O_{7-\delta}$. In addition, the achievement of optimal $T_c^{onset} = 78.2$ K and $T_c^{zero} = 40$ K at 15 GPa for $La_2PrNi_2O_{7-\delta}$ polycrystalline samples surpasses the corresponding values of the undoped $La_3Ni_2O_7$ at the samiliar pressure, which is consistent with the results of band structure calculation.

## Conclusion

In summary, a series of La$_{3-x}$Pr$_x$Ni$_2$O$_{7-\delta}$ ($x$=0, 0.1, 0.3, 1) polycrystalline samples were synthesized using the sol-gel method, and the evolution of their structure, magnetic and electrical transport properties as a function of doping concentration were investigated at ambient pressure. Our results show that all these obtained La$_{3-x}$Pr$_x$Ni$_2$O$_{7-\delta}$ (0.0≤$x$≤1.0) samples maintain the same orthorhombic *Amam* structure as undoped La$_3$Ni$_2$O$_7$ at ambient conditions. In addition, the lattice parameters display a negative correlation with the Pr-doping concentration, confirming the successful introduction of chemical pressure. *In situ* high-pressure structural analysis reveals that the chemically precompressed La$_2$PrNi$_2$O$_{7-\delta}$ sample undergoes a structure transition from the orthorhombic *Amam* space group to the tetragonal *I*4/*mmm* space group under high pressure at 10-11 GPa, which is different with that observed in undoped La$_3$Ni$_2$O$_7$. Remarkably, the high-pressure tetragonal phase of La$_2$PrNi$_2$O$_{7-\delta}$ shares the same symmetry as many cuprate superconductors and was surprisingly found to show high-temperature superconductivity with a stronger pressure coefficient of d$T_c$/d$P$ in comparison with that of undoped La$_3$Ni$_2$O$_7$. The discovery of high-temperature superconductivity in the high-pressure tetragonal phase of La$_2$PrNi$_2$O$_{7-\delta}$ broadens the nickelate superconducting family and provides a new platform to investigate underlying mechanisms.

## Acknowledgments

This work is supported by the National Natural Science Foundation of China (12025408, 11921004, 11888101), National Key Research and Development Program of China (2021YFA1400200), the Strategic Priority Research Program of CAS (XDB33000000), the Specific Research Assistant Funding Program of CAS (E3VP011X61). The high-pressure experiments were performed at the Cubic Anvil Cell station of Synergic Extreme Condition User Facility (SECUF). High-pressure synchrotron X-ray measurements were performed at 4W2 High Pressure Station, Beijing Synchrotron Radiation Facility (BSRF), which is supported by Chinese Academy of Sciences.

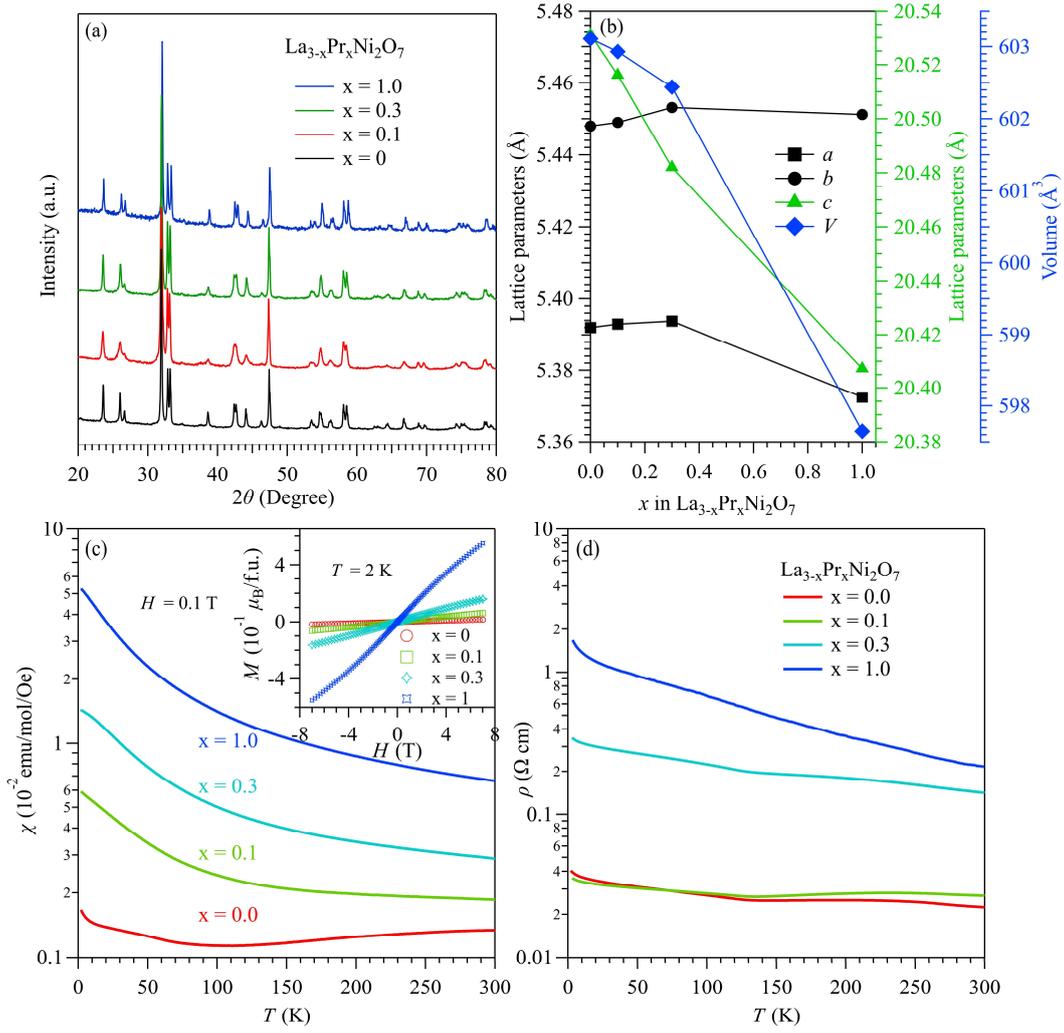

**Figure 1. Characterizations on the structure, electrical transport and magnetic properties of La$_{3-x}$Pr$_x$Ni$_2$O$_{7-\delta}$ (0.0≤*x*≤1.0) samples.** (a) The XRD patterns. (b) The obtained unit-cell parameters as a function of Pr content *x* . (c) Temperature dependence of magnetic susceptibility $\chi(T)$. The inset shows their isothermal magnetization *M*(*H*) curves measured between +7 T and -7 T at 2 K. (d) Temperature dependence of resistivity $\rho(T)$.

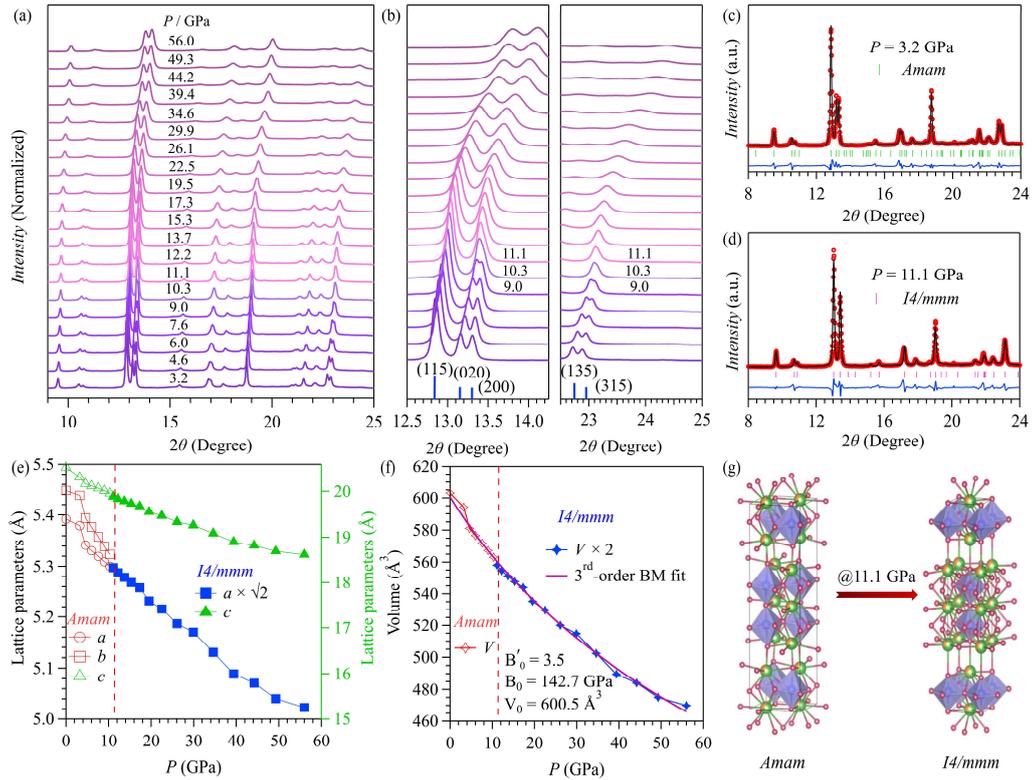

**Figure 2. In situ structural characterizations of La$_2$PrNi$_2$O$_{7-\delta}$ under high pressures.** (a) Synchrotron X-ray diffraction patterns of La$_2$PrNi$_2$O$_{7-\delta}$ powder samples under various pressure between 3.2 and 56 GPa. (b) The enlarged view of HP-SXRD around the representative 2θ range, highlighting the gradual mergeing of the diffraction peaks upon compression. (c) and (d) Refinement results of the synchrotron X-ray diffraction patterns at 3.2 GPa using the space group *Amam* and 11.1 GPa using the space group *I4/mmm*. The experimental data and fitted XRD profile were shown as red circles and black lines, respectively. The green and pink bars show the positions of the calculated Bragg reflections for *Amam* and *I4/mmm* phases. The difference between the observed and the fitted XRD patterns was shown with a blue line at the bottom of the diffraction peaks. (e) Lattice parameters and (f) Cell volume as a function of the pressure, which can be fitted by using the third-order Birch-Murnaghan equation. (g) Crystal structure transformation of La$_2$PrNi$_2$O$_{7-\delta}$ under high pressure.

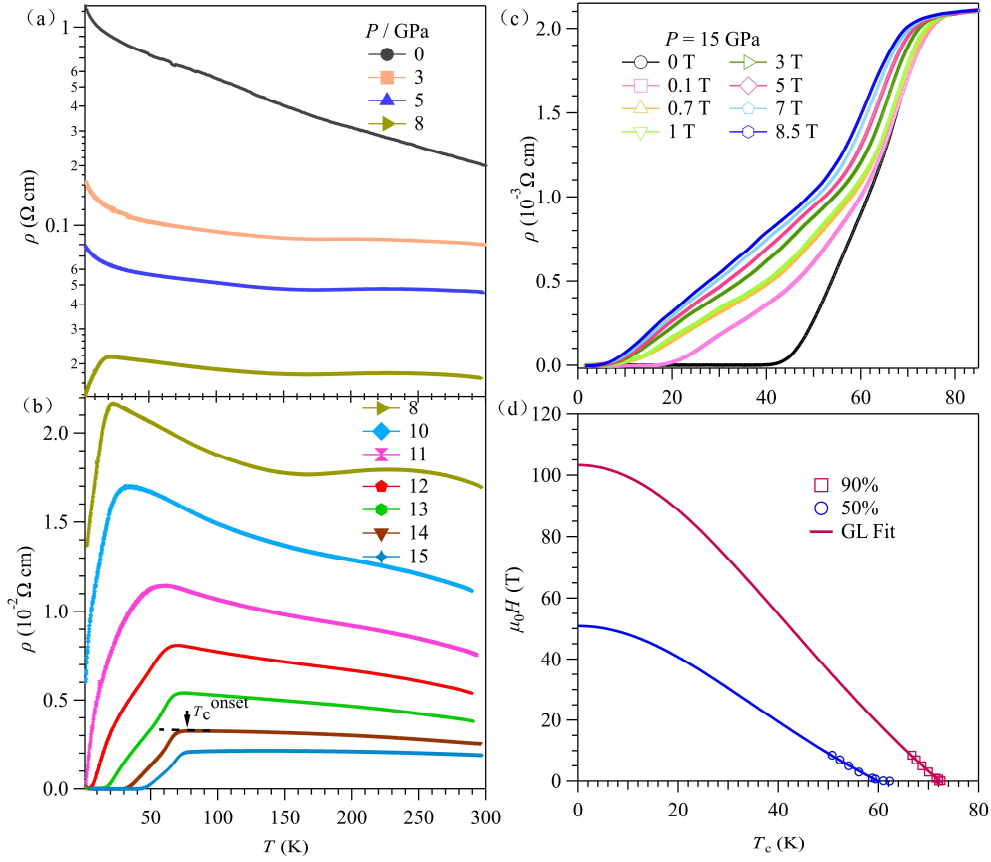

**Figure 3. Superconducting transition in $La_2PrNi_2O_{7-\delta}$ under high pressure.** (a) and (b) Temperature dependence of resistivity $\rho(T)$ of the $La_2PrNi_2O_{7-\delta}$ sample under various hydrostatic pressures up to 15 GPa measured in a CAC employing glycerol as the liquid pressure transmitting medium. Here, the $T_c^{onset}$ is determined as the interception between two straight lines below and above the superconducting transitions. (c) The low-temperature resistivity $\rho(T)$ at 15 GPa under various magnetic fields up to 8.5 T. (d) Temperature dependence of the upper critical field $\mu_0H_{c2}(T)$ for the $La_2PrNi_2O_{7-\delta}$ sample at 15 GPa. The solid line is the fitting curve by using the formula $H_{c2}= H_{c2}(0)(1-t^2)/(1+t^2)$, where $t = T/T_c$.

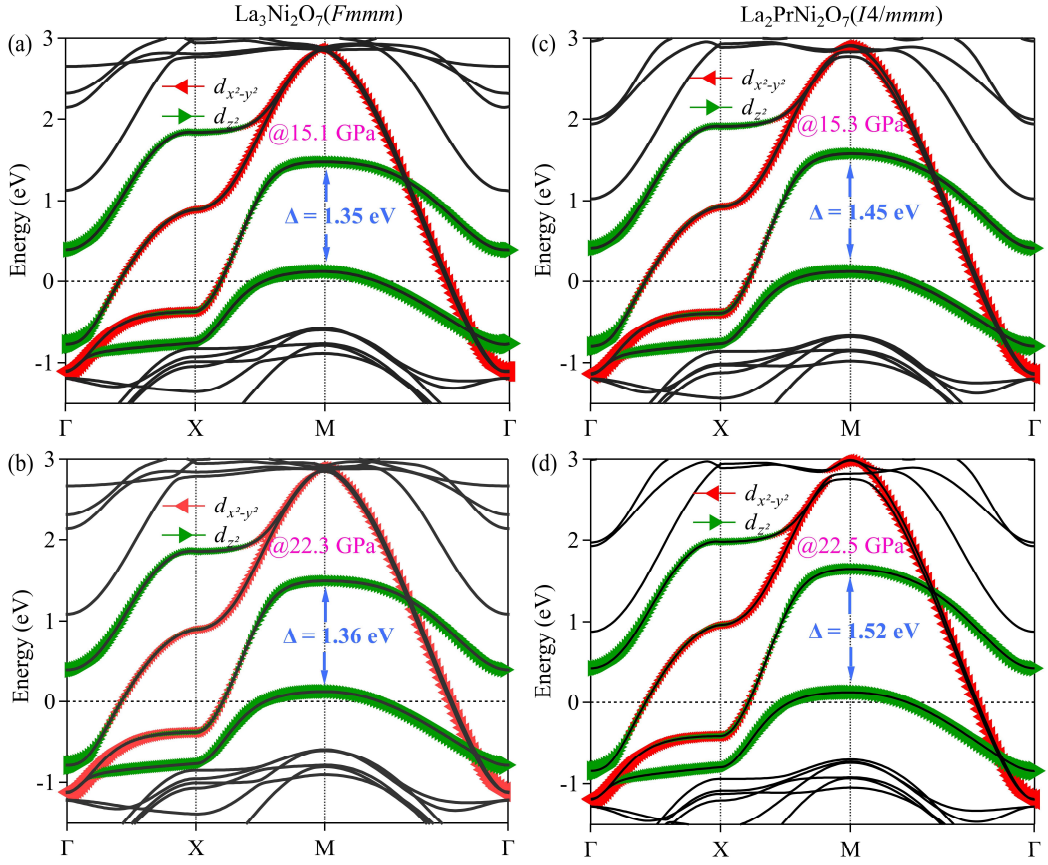

**Fig 4. Comparision of projected band structures of La$_3$Ni$_2$O$_7$ and La$_2$PrNi$_2$O$_7$ under high pressures.** (a) 15.1GPa and (b) 22.3 GPa for the HP *Fmmm* phase of La$_3$Ni$_2$O$_7$. (c) 15.3GPa and (d) 22.5 GPa for the HP *I4/mmm* phase of La$_2$PrNi$_2$O$_7$. The weight of different Ni orbitals are represented by the size of the triangles.

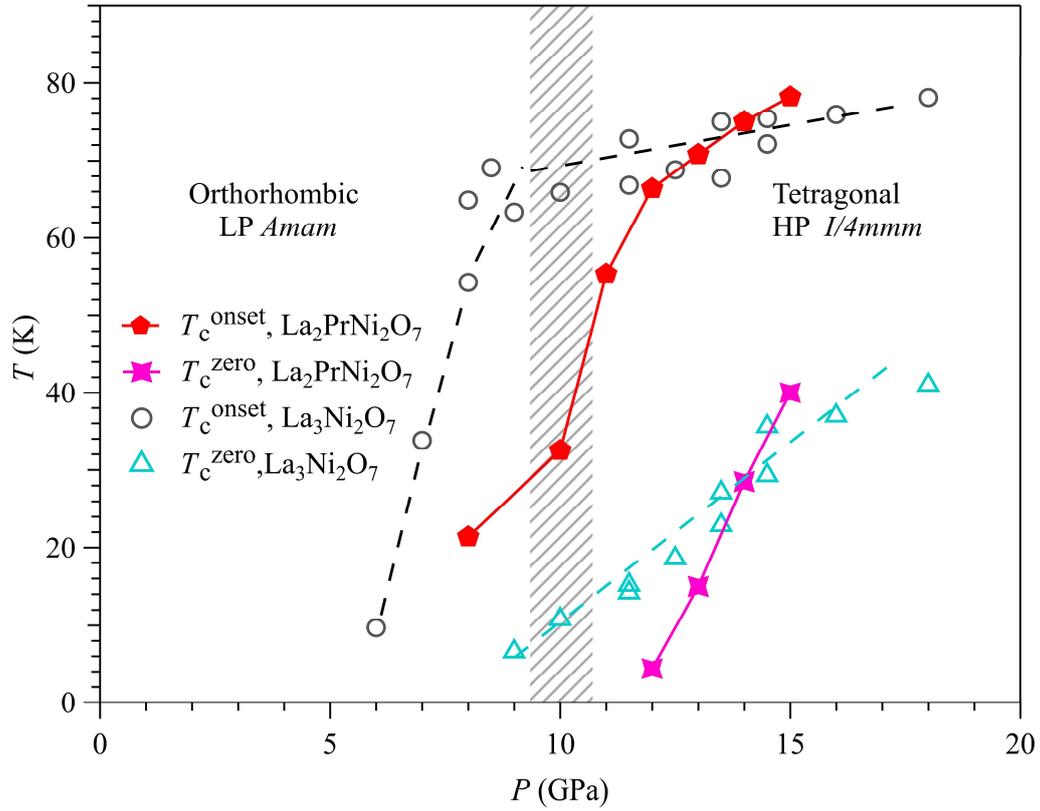

**Fig 5. Phase diagram of the high-temperature superconductivity in the La$_2$PrNi$_2$O$_{7-\delta}$ sample in comparison with that of La$_3$Ni$_2$O$_7$.** The filled marks represent the critical superconducting transition temperatures of La$_2$PrNi$_2$O$_{7-\delta}$ determined from the electrical transport measurements in CAC. The open marks for La$_3$Ni$_2$O$_7$ are taken from Ref. [43]

# Supplementary Information

# Observation of high-temperature superconductivity in the high-pressure tetragonal phase of La$_2$PrNi$_2$O$_{7-\delta}$

G. Wang[1,2#], N. N. Wang[1,2#*], Y. X. Wang[1,2#], L. F. Shi[1,2], X. L. Shen[3], J. Hou[1,2], H. M. Ma[3], P. T. Yang[1,2], Z. Y. Liu[1,2], H. Zhang[1,2], X. L. Dong[1,2], J. P. Sun[1,2], B. S. Wang[1,2], K. Jiang[1,2*], J. P. Hu[1,2], Y. Uwatoko[3], and J.-G. Cheng[1,2*]

[1]*Beijing National Laboratory for Condensed Matter Physics and Institute of Physics, Chinese Academy of Sciences, Beijing 100190, China*

[2]*School of Physical Sciences, University of Chinese Academy of Sciences, Beijing 100190, China*

[3]*Institute for Solid State Physics, University of Tokyo, Kashiwa, Chiba 277-8581, Japan*

\# These authors contribute equally to this work.

*Corresponding authors: nnwang@iphy.ac.cn; jiangkun@iphy.ac.cn; jgcheng@iphy.ac.cn

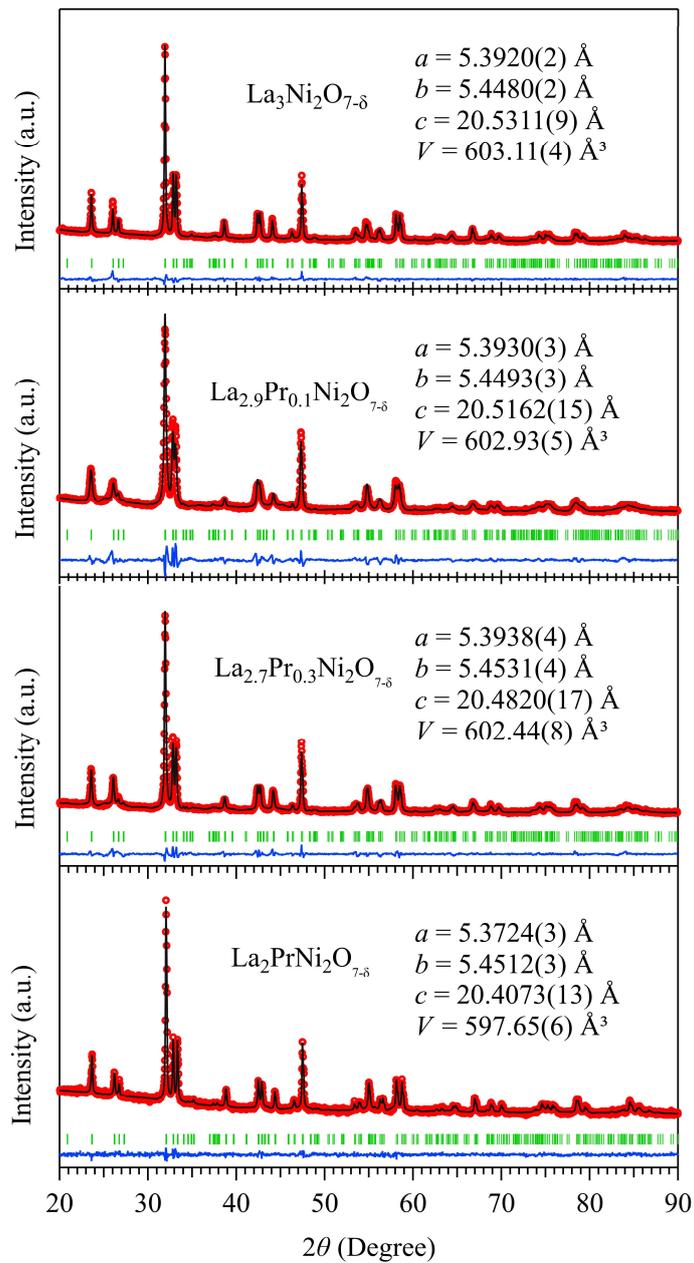

Fig S1. Rietveld refinement XRD patterns and lattice parameters of $La_{3-x}Pr_xNi_2O_{7-\delta}$ (x = 0, 0.1, 0.3 and 1). The green bars and blue lines at the bottom correspond to the calculated Bragg diffraction positions and the discrepancies between observed and calculated values.

Table S1. The crystal parameters calculated by DFT relaxation for the low-pressure (LP) and high-pressure (HP) phase of $La_2PrNi_2O_7$. $d_{apical}$ is the distance between Ni and apical O, $d_{planar}$ is the distance between Ni and the planar O. This table shows that for $La_2PrNi_2O_7$, the LP phase prefer to be the orthorhombic phase while the HP phase prefer to be the tetragonal phase even under a virtual 0 GPa external pressure. Another notable trend is that, with increasing pressure, whether in the LP or HP range, the difference between lattice constants *a* and *b* decreases, indicating a tendency toward a tetragonal phase under high pressure.

**LP *Amam* phase of $La_2PrNi_2O_7$:**

| pressure | *a* (Å) | *b* (Å) | *c* (Å) | dapical (Å) | dplanar (Å) | crystal system |
|---|---|---|---|---|---|---|
| 0 Gpa | 5.4010 | 5.5278 | 20.1628 | 2.197 | 1.953 | orthorhombic |
| 5 Gpa | 5.3567 | 5.4613 | 19.9397 | 2.15 | 1.928 | orthorhombic |
| 10 Gpa | 5.3193 | 5.3949 | 19.7706 | 2.116 | 1.905 | orthorhombic |

**HP *I4/mmm* phase of $La_2PrNi_2O_7$:**

| pressure | *a* (Å) | *b* (Å) | *c* (Å) | dapical (Å) | dplanar (Å) | crystal system |
|---|---|---|---|---|---|---|
| 0 Gpa | 5.4262 | 5.4226 | 20.1930 | 2.168 | 1.918 | nearly tetragonal |
| 15 Gpa | 5.2972 | 5.2974 | 19.6739 | 2.09 | 1.874 | tetragonal |
| 20 Gpa | 5.2606 | 5.2607 | 19.5366 | 2.071 | 1.861 | tetragonal |
| 30 Gpa | 5.1966 | 5.1966 | 19.2974 | 2.038 | 1.839 | tetragonal |